\documentclass[12pt]{article}
\usepackage{amsmath}
\usepackage{amssymb}
\usepackage{amsfonts}
\usepackage{amscd}
\usepackage[all]{xy}

\title{Ladder operators for solvable potentials connected with exceptional orthogonal polynomials}

\author{C. Quesne\\
{\small \sl Physique Nucl\'eaire Th\'eorique et Physique Math\'ematique, Universit\'e Libre de Bruxelles,} \\
{\small \sl Campus de la Plaine CP229, Boulevard du Triomphe, B-1050 Brussels, Belgium,}\\
{\small \sl E-mail: cquesne@ulb.ac.be}}

\date{ }
\begin{document}
\baselineskip=22pt plus 1pt minus 1pt
\maketitle

\begin{abstract}
Exceptional orthogonal polynomials constitute the main part of the bound-state wavefunctions of some solvable quantum potentials, which are rational extensions of well-known shape-invariant ones. The former potentials are most easily built from the latter by using higher-order supersymmetric quantum mechanics (SUSYQM) or Darboux method. They may in general belong to three different types (or a mixture of them): types I and II, which are strictly isospectral, and type III, for which k extra bound states are created below the starting potential spectrum. A well-known SUSYQM method enables one to construct ladder operators for the extended potentials by combining the supercharges with the ladder operators of the starting potential. The resulting ladder operators close a polynomial Heisenberg algebra (PHA) with the corresponding Hamiltonian. In the special case of type III extended potentials, for this PHA the k extra bound states form k singlets isolated from the higher excited states. Some alternative constructions of ladder operators are reviewed. Among them, there is one that combines the state-adding and state-deleting approaches to type III extended potentials (or so-called Darboux-Crum and Krein-Adler transformations) and mixes the k extra bound states with the higher excited states. This novel approach can be used for building integrals of motion for two-dimensional superintegrable systems constructed from rationally-extended potentials.
\end{abstract}

%
%
\newpage
\section{Introduction}

In 2008, the introduction of exceptional orthogonal polynomials (EOP) by G\'omez-Ullate, Kamran, and Milson \cite{gomez10a, gomez09} has constituted a big advance in both mathematics and physics. They indeed form complete and orthogonal polynomial systems generalizing the classical orthogonal polynomials (COP) of Hermite, Laguerre, and Jacobi. In contrast with the latter, there are indeed some gaps in the sequence of their degrees, which does not impede them to form complete systems. Furthermore, they occur in bound-state wavefunctions of some supersymmetric partners of shape-invariant potentials, which are rational extensions of the latter \cite{cq08}.\par
%
%
Since then, a lot of work has been carried out in this area (see, e.g., \cite{gomez10b, gomez12a, gomez12b, gomez13, gomez14a, gomez14b, fellows, bagchi09, cq09, cq11a, cq11b, cq12a, cq12b, marquette13a, odake09, odake10, sasaki, odake11, odake13a, odake13b, grandati11a, grandati11b, grandati12a, grandati12b, grandati13, ho11a, ho11b}). It turned out that the easiest approach to construct solvable potentials connected with EOP is $n$th-order supersymmetric quantum mechanics (SUSYQM) or, equivalently, some extensions of the Darboux method \cite{andrianov}. In such an approach, one starts from $n$ different seed solutions $\varphi_1, \varphi_2, \ldots, \varphi_n$ of a starting Hamiltonian $H^{(1)}$ and one gets the potential of the partner as $V^{(2)}(x) = V^{(1)}(x) - 2 \frac{d^2}{dx^2} {\cal W}(\varphi_1, \varphi_2, \ldots, \varphi_n)$, provided the Wronskian $\cal W$ is nonsingular.\par
%
%
In general, there may occur three types of EOP (I, II, and III). For I and II, $H^{(2)}$ is strictly isospectral to $H^{(1)}$ and shape invariant. For III, there are some additional levels below the spectrum of $H^{(1)}$ and $H^{(2)}$ is not shape invariant. Note that for the harmonic oscillator, only type III EOP occur.\par
%
%
The purpose of the present communication is to review some recent works, made in collaboration with I. Marquette, on the construction of ladder operators for the potentials connected with EOP \cite{marquette13a, marquette13b, marquette13c, marquette14}.\par
%
%
\section{Ladder operators in SUSYQM}

In $n$th-order SUSYQM, let us consider a set of two partner Hamiltonians
\begin{equation}
  H^{(i)} = - \frac{d^2}{dx^2} + V^{(i)}(x), \qquad i=1, 2,
\end{equation}
and let us assume that $H^{(1)}$ has some ladder operators $a$ and $a^{\dagger}$ of order $k$. These generate with $H^{(1)}$ a polynomial Heisenberg algebra (PHA) of order $k-1$ \cite{fernandez99, fernandez05, carballo}, whose commutation relations are given by
\begin{equation}
\begin{split}
  & [H^{(1)}, a^{\dagger}] = \lambda a^{\dagger}, \qquad [H^{(1)}, a] = - \lambda a, \\
  & [a, a^{\dagger}] = P^{(1)}(H^{(1)}+ \lambda) - P^{(1)}(H^{(1)}), 
\end{split}  
\end{equation}
where $\lambda$ is some constant and $P^{(1)}(H^{(1)})$ is some $k$th-degree polynomial in $H^{(1)}$, $P^{(1)}(H^{(1)}) = \prod_{i=1}^k (H^{(1)} - \epsilon_i)$.\par
%
%
Since $a$ and/or $a^{\dagger}$ may have zero modes, this PHA may have infinite-dimensional unitary irreducible representations (unirreps), as well as finite-dimensional ones (singlet, doublet, and more generally multiplet).\par
%
%
Well-known examples are those of the harmonic oscillator and of the radial harmonic oscillator. In the former case, $a$ and $a^{\dagger}$ may be written as
\begin{equation}
  a = \frac{d}{dx} + x, \qquad a^{\dagger} = - \frac{d}{dx} + x,
\end{equation}
and we have $k=1$, $\lambda=2$ and $P^{(1)}(H^{(1)}) = H^{(1)} - 1$. The corresponding PHA then reduces to the Heisenberg algebra. In the latter case, we may write
\begin{equation}
\begin{split}
  & a = \frac{1}{4} \left(2 \frac{d^2}{dx^2} + 2x \frac{d}{dx} + \frac{1}{2} x^2 - \frac{2\ell(\ell+1)}{x^2} + 
  1\right), \\
  & a^{\dagger} = \frac{1}{4} \left(2 \frac{d^2}{dx^2} - 2x \frac{d}{dx} + \frac{1}{2} x^2 - 
  \frac{2\ell(\ell+1)}{x^2} - 1\right),
\end{split}
\end{equation}
where $\ell$ denotes the angular momentum quantum number. Now $k=2$, $\lambda=2$, $P^{(1)}(H^{(1)}) = \frac{1}{16}(2H^{(1)} - 3 - 2\ell) (2H^{(1)} - 1 + 2\ell)$ and the resulting PHA is the su(1,1) Lie algebra. In more general cases, the PHA turns out to be nonlinear.\par
%
%
In $n$th-order SUSYQM, the two Hamiltonians $H^{(1)}$ and $H^{(2)}$ intertwine with $n$th-order differential operators $\cal A$ and ${\cal A}^{\dagger}$ as ${\cal A} H^{(1)} = H^{(2)} {\cal A}$, ${\cal A}^{\dagger} H^{(2)} = H^{(1)} {\cal A}^{\dagger}$. The operators $\cal A$ and ${\cal A}^{\dagger}$, which may be written in terms of the $n$ seed solutions $\varphi_1, \varphi_2, \ldots, \varphi_n$ of $H^{(1)}$, enable one to relate the energy spectra, the wavefunctions and the ladder operators of $H^{(1)}$ and $H^{(2)}$ \cite{andrianov, fernandez05}.\par
%
%
In particular, from the ladder operators $a$ and $a^{\dagger}$ of $H^{(1)}$, one gets ladder operators $b$ and $b^{\dagger}$ of $H^{(2)}$, which are $(2n+k))$th-order differential operators given by $b = {\cal A} a {\cal A}^{\dagger}$ and $b^{\dagger} = {\cal A} a^{\dagger} {\cal A}^{\dagger}$, respectively \cite{fernandez99, fernandez05}. The new operators $H^{(2)}$, $b^{\dagger}$ and $b$ satisfy another PHA, which is entirely determined by the first one and the SUSYQM transformation. Their commutation relations are indeed given by
\begin{equation}
\begin{split}
  & [H^{(2)}, b^{\dagger}] = \lambda b^{\dagger}, \qquad [H^{(2)}, b] = - \lambda b, \\
  & [b, b^{\dagger}] = P^{(2)}(H^{(2)}+ \lambda) - P^{(2)}(H^{(2)}), 
\end{split}  
\end{equation}
where $P^{(2)}(H^{(2)})$ is the $(2n+k)$th-degree polynomial in $H^{(2)}$ defined by
\begin{equation}
\begin{split}
  & P^{(2)}(H^{(2)}) = P^{(1)}(H^{(2)}) f(H^{(2)}-\lambda) f(H^{(2)}), \\
  & f(H^{(2)}) = {\cal A} {\cal A}^{\dagger}.
\end{split}
\end{equation}
\par
%
%
\section{\boldmath Ladder operators for EOP-related problems}

In Ref.~\cite{marquette13b}, we used the method presented in Sect.~2 to construct ladder operators for potentials related to (type III) Hermite and (type I, II, or III) Laguerre EOP.\par
%
%
{}For type I or II, corresponding to isospectrality, the properties of $(H^{(1)}, a, a^{\dagger})$ are transferred to $(H^{(2)}, b, b^{\dagger})$ and $b$, $b^{\dagger}$ act on all the wavefunctions of $H^{(2)}$. In contrast, for type III, although the properties of $(H^{(1)}, a, a^{\dagger})$ are still transferred to $(H^{(2)}, b, b^{\dagger})$, the operators $b$, $b^{\dagger}$ do not see the states that have been added below the spectrum of $H^{(1)}$. Hence, apart from some unirreps similar to those present for $H^{(1)}$, there are also $n$ singlets corresponding to the added states.\par
%
%
When one constructs higher-dimensional superintegrable systems whose constituent Hamiltonians are connected with EOP families, ladder operators of these one-dimensional Hamiltonians are very useful to build integrals of the motion of order higher than two by using a well-known prescription \cite{marquette10}. However, in order to be able to algebraically derive the energy spectrum of the superintegrable systems, together with the degeneracies of their levels, from the representations of the polynomial algebra generated by the integrals of motion, it is necessary that such ladder operators act on all eigenstates of the constituent Hamiltonians. In Ref.~\cite{marquette13b}, this condition was only satisfied for the type I or II case, but not for type III.\par
%
%
In a first attempt to deal with this problem \cite{marquette13a}, we proposed some new ladder operators, different from $b$ and $b^{\dagger}$, for rational extensions of the harmonic oscillator, associated with type III Hermite EOP and constructed in second-order SUSYQM. It turned out, however, that the two added states below the spectrum of the oscillator formed a doublet and were not connected to higher excited states.\par
%
%
In a second attempt, we succeeded in solving the problem in the cases of rational extensions of the harmonic oscillator and of the radial harmonic oscillator, constructed in $n$th-order SUSYQM \cite{marquette13c, marquette14}. The procedure used will be reviewed in the next section.\par    
%
%
\section{\boldmath State adding versus state deleting in $n$th-order SUSYQM}

In general, there are several possibilities for choosing the $n$ different seed solutions $\varphi_1, \varphi_2, \ldots, \varphi_n$ of the starting Hamiltonian $H^{(1)}$ and these go back to the Darboux intertwining method \cite{darboux}.\par
%
%
Broadly speaking, there are two (equivalent) possibilities: that of adding a number of states below the spectrum of the starting Hamiltonian, which is the method alluded to above, and that of deleting a number of  bound states in this spectrum. The former is referred to as the Darboux-Crum method and takes its origin in the Crum extension \cite{crum} of the Darboux method. In such a case, the seed functions are unphysical solutions of the Schr\"odinger equation for $H^{(1)}$ with energy less than the ground-state, which are converted into physical solutions of $H^{(2)}$. The latter is called the Krein-Adler method and comes from a generalization of the Crum method made independently by Krein \cite{krein} and Adler \cite{adler}. The seed functions are then chains of bound-state wavefunctions that may be lacunary with some even gaps. Note that in general the two $n$ values are different in both approaches.\par
%
%
To distinguish both possibilities, we shall use different notations, as follows:

\begin{itemize}

\item[1)] $(H^{(1)}, H^{(2)})$ with $H^{(i)} = - \frac{d^2}{dx^2} + V^{(i)}(x)$, $i=1,2$, in the case of Darboux-Crum method;
\vskip 0.2cm
\item[2)] $(\bar{H}^{(1)}, \bar{H}^{(2)})$ with $\bar{H}^{(i)} = - \frac{d^2}{dx^2} + \bar{V}^{(i)}(x)$, $ i=1,2$, in the case of Krein-Adler method.

\end{itemize}
\par
%
%
We plan to show that in the cases of the harmonic oscillator and of the radial harmonic oscillator, it is possible to arrive at $\bar{V}^{(2)}(x) = V^{(2)}(x) + \mbox{\rm constant}$, thereby allowing the construction of ladder operators in a novel way \cite{marquette14}.\par 
%
%
\section{Ladder operators for rational extensions of the harmonic oscillator}

In this case, we start from the same harmonic oscillator Hamiltonian, $\bar{H}^{(1)} = H^{(1)}$, which means that $\bar{V}^{(1)}(x) = V^{(1)}(x) = x^2$, $-\infty < x < \infty$.\par
%
%
In the state-adding case, we take $n \to k$ and
\begin{equation}
  \varphi_i(x) \to \phi_{m_i}(x) = {\cal H}_{m_i}(x) e^{x^2/2}, \qquad {\cal H}_{m_i}(x) = (-{\rm i})^{m_i} 
  H_{m_i}({\rm i}x), \qquad i=1, 2, \ldots, k, 
\end{equation}
where $H_m(x)$ denotes a Hermite polynomial. The $\phi_{m_i}(x)$'s are eigenfunctions of $H^{(1)}$, corresponding to eigenvalues $E_{m_i} = - 2m_i -1$ below the ground state. For even $m$ values, they are nodeless on the whole real line, while for odd $m$ ones, they have a single zero at $x=0$. With the choice $m_1 < m_2 < \cdots < m_k$ with $m_i$ even (resp.\ odd) for $i$ odd (resp.\ even), the partner $V^{(2)}(x)$ turns out to be nonsingular \cite{fernandez05}. On using standard properties of Wronskians \cite{muir}, one can then show that $V^{(2)}(x)$ becomes
\begin{equation}
  V^{(2)}(x) = x^2 - 2k - 2 \frac{d^2}{dx^2} \log {\cal W}({\cal H}_{m_1}, {\cal H}_{m_2}, \ldots, 
  {\cal H}_{m_k})  \label{eq:V-2}
\end{equation}
with corresponding spectrum 
\begin{equation}
  E^{(2)}_{\nu} = 2\nu+1, \qquad \nu = -m_k-1, \ldots, -m_2-1, -m_1-1, 0, 1, 2, \ldots. \label{eq:partner-1}
\end{equation}
\par
%
%
On the other hand, in the state-deleting case, we take $n \to m_k-k+1$ and
\begin{equation}
  (\varphi_1, \varphi_2, \ldots, \varphi_n) \to (\psi_1, \psi_2, \ldots, \check{\psi}_{m_k-m_{k-1}}, \ldots, 
     \check{\psi}_{m_k-m_2}, \ldots, \check{\psi}_{m_k-m_1}, \ldots, \psi_{m_k}),  \label{eq:deleted-wf}
\end{equation}
where $\psi_{\nu}$ denotes a bound-state wavefunction and $\check{\psi}_{\nu}$ means that $\psi_{\nu}$ is excluded from the list. The wavefunctions in (\ref{eq:deleted-wf}) are then suppressed from the spectrum. Standard properties of Wronskians lead again to
\begin{equation}
  \bar{V}^{(2)}(x) = x^2 + 2(m_k+1-k) - 2 \frac{d^2}{dx^2} \log {\cal W}(H_1, H_2, \ldots, 
  \check{H}_{m_k-m_{k-1}}, \ldots, \check{H}_{m_k-m_1}, \ldots, H_{m_k}),  \label{eq:V-bar-2}
\end{equation}
which is nonsingular provided the gaps between the surviving levels with $\nu = 0, m_k-m_{k-1}, \ldots, m_k-m_2, m_k-m_1, m_k+1, m_k+2, \ldots$, correspond to even numbers of consecutive levels \cite{krein, adler}. This leads to the same conditions on $m_1$, $m_2$, \ldots, $m_k$ as in the state-adding case. After redefining $\nu$, one gets the spectrum 
\begin{equation}
  \bar{E}^{(2)}_{\nu} = 2m_k+2\nu+3, \qquad \nu = -m_k-1, \ldots, -m_2-1, -m_1-1, 0, 1, 2, \ldots.
  \label{eq:partner-2}
\end{equation}
\par 
%
%
As it has been proved that the two Wronskians in (\ref{eq:V-2}) and (\ref{eq:V-bar-2}) only differ by some multiplicative constant \cite{odake13a}, the two obtained potentials $V^{(2)}(x)$ and $\bar{V}^{(2)}(x)$ are the same up to some additive constant, 
\begin{equation}
  \bar{V}^{(2)}(x) = V^{(2)}(x) + 2m_k + 2.
\end{equation}
This result agrees with the corresponding energies (\ref{eq:partner-1}) and (\ref{eq:partner-2}).
\par
%
%
SUSYQM intertwining relations show that one can go from $H^{(2)}$ to $H^{(2)} + 2m_k + 2$ along the following path
\begin{equation}
  \xymatrixcolsep{5pc}\xymatrix@1{
  H^{(2)}  \ar@/_{10mm}/[rr]^{c}  \ar[r]^{{\cal A}^{\dagger}} & H^{(1)} = \bar{H}^{(1)}  
  \ar[r]^-{\bar{\cal A}} & \bar{H}^{(2)} = H^{(2)}+2m_k+2}  
\end{equation}
The $(m_k+1)$th-order differential operator
\begin{equation}
  c = \bar{\cal A} {\cal A}^{\dagger}
\end{equation}
that performs such a transformation is therefore a lowering operator for $H^{(2)}$.\par
%
%
Together with its Hermitian conjugate and $H^{(2)}$, it satisfies a PHA of $m_k$th order, defined by the commutation relations
\begin{equation}
\begin{split}
  & [H^{(2)}, c^{\dagger}] = (2m_k+2) c^{\dagger}, \qquad [H^{(2)}, c] = - (2m_k+2) c, \\
  & [c, c^{\dagger}] = Q(H^{(2)}+2m_k+2) - Q(H^{(2)}), \\
  & Q(H^{(2)}) = \left(\prod_{i=1}^k (H^{(2)} + 2m_i + 1)\right) 
         \left(\prod_{\substack{j=1 \\ j \ne m_k-m_{k-1}, \ldots, m_k-m_1}}^{m_k} (H^{(2)} - 2j -1)\right). 
\end{split}  \label{eq:PHA}  
\end{equation}
As explained in \cite{marquette14}, the explicit form of $Q(H^{(2)})$, given in (\ref{eq:PHA}), can be obtained from standard properties of SUSYQM.
\par
%
%
{}From the action of $Q(H^{(2)}) = c^{\dagger} c$ on $\psi^{(2)}_{\nu}(x)$ obtained by replacing $H^{(2)}$ by $E^{(2)}_{\nu}$, that of $c$ and $c^{\dagger}$ on the same can be easily derived. By choosing the normalization constants of the $\psi^{(2)}_{\nu}$'s in such a way that all matrix elements of the ladder operators $c^{\dagger}$, $c$ are nonnegative, one can write the results for $c$ as
\begin{equation}
\begin{split}
  & c\psi^{(2)}_{\nu} = 0, \qquad \nu=-m_k-1, \ldots, -m_1-1, 1, 2, \ldots, m_k-m_{k-1}-1,  \\
  & \qquad  m_k-m_{k-1}+1, \ldots, m_k-m_1-1, m_k-m_1+1, \ldots, m_k,  \label{eq:action-0}
\end{split}
\end{equation}
\begin{equation}
  c\psi^{(2)}_0 = \Biggl[2^{m_k+1} (m_k+1)! \Biggl(\prod_{i=1}^{k-1} \frac{m_i+1}{m_k-m_i}\Biggr)\Biggr]^{1/2} 
  \psi^{(2)}_{-m_k-1},  \label{eq:action-1} 
\end{equation}
\begin{equation}
\begin{split}
  & c\psi^{(2)}_{m_k-m_i} =  \Biggl[2^{m_k+1} (m_k+1) (2m_k-m_i+1) (m_k-m_i-1)! m_i! \\
  & \quad \times \Biggl(\prod_{j=1}^{i-1} \frac{m_k+m_j-m_i+1}{m_i-m_j}\Biggr) \Biggl(\prod_{l=i+1}^{k-1} 
       \frac{m_k+m_l-m_i+1}{m_l-m_i}\Biggr)\Biggr]^{1/2} \psi^{(2)}_{-m_i-1}, \\
  & \quad \qquad i=1, 2, \ldots, k-1, 
\end{split}
\end{equation}
\begin{equation}
\begin{split}
  & c\psi^{(2)}_{\nu} =  \Biggl[2^{m_k+1} (\nu+m_k+1) \frac{(\nu-1)!}{(\nu-m_k-1)!} \Biggl(\prod_{i=1}^{k-1} \frac{\nu+m_i+1}
        {\nu+m_i-m_k}\Biggr)\Biggr]^{1/2} \psi^{(2)}_{\nu-m_k-1}, \\
  & \quad \qquad \nu=m_k+1, m_k+2, \ldots,
\end{split}  \label{eq:action-2}
\end{equation}   
those for $c^{\dagger}$ being deduced by using Hermitian conjugation. This proves that the above PHA has $m_k+1$ infinite-dimensional unirreps, whose lowest-weight states are $\psi^{(2)}_i(x)$ with $i=-m_k-1$, \ldots, $-m_1-1$, 1, 2, \ldots, $m_k-m_{k-1}-1$, $m_k-m_{k-1}+1$, \ldots, $m_k-m_1-1$, $m_k-m_1+1$, \ldots, $m_k$, respectively. Hence, the $k$ lowest states are mixed with the higher ones.\par  
%
%
\section{Ladder operators for rational extensions of the radial harmonic oscillator}

The radial harmonic oscillator case is more complicated because SUSYQM changes the $\ell$ value in the potential $V_{\ell}(x) = \frac{1}{4} x^2 + \frac{\ell(\ell+1)}{x^2}$, $0<x<\infty$. We sketch it below and refer the reader to \cite{marquette14} for more details.\par
%
%
In the state-adding case, we take $n \to k$, $V^{(1)}(x) = V_{\ell + k}(x)$, and
\begin{equation}
  \varphi_i(x) \to \phi^{(\ell+k)}_{m_i}(x) \propto x^{-\ell-k} e^{\frac{1}{4}x^2} 
  L^{(-\ell-k-\frac{1}{2})}_{m_i}\left(-\frac{1}{2}x^2\right), \qquad i=1, 2, \ldots, k, 
\end{equation}
where $L^{(-\alpha-k)}_m(-z)$, with $z=\frac{1}{2}x^2$ and $\alpha=\ell+\frac{1}{2}$, denotes a Laguerre polynomial, $m_1 < m_2 < \cdots < m_k$, $m_i$ is even (resp.\ odd) for $i$ odd (resp.\ even), and $\alpha+k > m_k$. Then it turns out that $V^{(2)}(x)$ is a rationally-extended $V_{\ell}(x)$ potential,
\begin{equation}
  V^{(2)}(x) = V_\ell(x) - k - 2 \frac{d^2}{dx^2} \log \tilde{\cal W}\bigl(L^{(-\alpha-k)}_{m_1}(-z), 
  L^{(-\alpha-k)}_{m_2}(-z), \ldots, L^{(-\alpha-k)}_{m_k}(-z)\bigr),
\end{equation}
where $\tilde{\cal W}\bigl(f_1(z), f_2(z), \ldots, f_k(z)\bigr)$ denotes the Wronskian of the functions $f_1(z), f_2(z), \ldots, f_k(z)$ with respect to $z$,
and that 
\begin{equation}
  E^{(2)}_{\ell,\nu} = 2\nu+\ell+k+\frac{3}{2}, \qquad \nu = -m_k-1, \ldots, -m_2-1, -m_1-1, 0, 1, 2, \ldots. 
\end{equation}
\par
%
%
In the state-deleting case, we take $n \to m_k+1-k$, $\bar{V}^{(1)}(x) = V_{\ell+k-m_k-1}(x)$, where $\alpha+k$ is assumed greater than $m_k+1$. Then, with the choice
\begin{equation}
\begin{split}
  & (\varphi_1, \varphi_2, \ldots, \varphi_n) \to \bigl(\psi^{(\ell+k-m_k-1)}_1, \psi^{(\ell+k-m_k-1)}_2, \ldots, 
       \check{\psi}^{(\ell+k-m_k-1)}_{m_k-m_{k-1}}, \ldots, \check{\psi}^{(\ell+k-m_k-1)}_{m_k-m_2}, 
       \ldots, \\
  & \quad \check{\psi}^{(\ell+k-m_k-1)}_{m_k-m_1}, \ldots, \psi^{(\ell+k-m_k-1)}_{m_k}\bigr),
\end{split}
\end{equation}
we obtain that $\bar{V}^{(2)}(x)$ is a rationally-extended $V_{\ell}(x)$,
\begin{equation}
\begin{split}
  \bar{V}^{(2)}(x) &= V_\ell(x) + m_k + 1 - k - 2 \frac{d^2}{dx^2} \log \tilde{{\cal W}}\bigl(
     L^{(\alpha+k-m_k-1)}_1(z), L^{(\alpha+k-m_k-1)}_2(z), \ldots, \\
  & \qquad \check{L}^{(\alpha+k-m_k-1)}_{m_k-m_{k-1}}(z), \ldots,  
     \check{L}^{(\alpha+k-m_k-1)}_{m_k-m_{1}}(z), \ldots, L^{(\alpha+k-m_k-1)}_{m_k}(z)\bigr), 
\end{split}  
\end{equation}
and 
\begin{equation}
  \bar{E}^{(2)}_{\ell, \nu} = 2\nu + \ell + k + m_k + \frac{5}{2}, \qquad \nu=-m_k-1, \ldots, -m_2-1, -m_1-1,
   0, 1, 2, \ldots. 
\end{equation}
%
%
Here, $V^{(2)}(x)$ and $\bar{V}^{(2)}(x)$ differ by an additive constant,
\begin{equation}
  \bar{V}^{(2)}(x) = V^{(2)}(x) + m_k + 1,
\end{equation}
again.\par
%
%
Since $\bar{H}^{(1)} \ne H^{(1)}$, we need a third SUSYQM transformation relating them to be able to construct ladder operators. As shown in Ref.~\cite{marquette14}, such a transformation is of $(m_k+1)$th order and the corresponding supercharge operators $\tilde{\cal{A}}$ and $\tilde{\cal{A}}^{\dagger}$ can be written in terms of $m_k+1$ seed solutions of class II, i.e., $n \to m_k+1$ and
\begin{equation}
  \varphi_i(x) \to \tilde{\phi}^{(\ell+k)}_{i-1}(x), \quad \tilde{\phi}^{(\ell+k)}_i(x) \propto x^{-\ell -k} 
  e^{-\frac{1}{4} x^2} L^{(-\ell-k-\frac{1}{2})}_i\left(\frac{1}{2} x^2\right).
\end{equation}
The corresponding partner potentials now read
\begin{equation}
\begin{split}
  \tilde{V}^{(1)}(x) &= V^{(1)}(x) = V_{\ell +k}(x), \\
  \tilde{V}^{(2)}(x) &= \tilde{V}^{(1)}(x) - 2 \frac{d^2}{dx^2} \log {\cal W}\bigl(\tilde{\phi}^{(\ell+k)}_0(x),
      \tilde{\phi}^{(\ell+k)}_1(x), \ldots, \tilde{\phi}^{(\ell+k)}_{m_k}(x)\bigr), 
\end{split}
\end{equation}
where the latter is nonsingular provided $\ell + k +1/2 > m_k$. The relation ${\cal W}\bigl(\tilde{\phi}^{(\ell+k)}_0, \tilde{\phi}^{(\ell+k)}_1, \ldots, \tilde{\phi}^{(\ell+k)}_{m_k}\bigr) \propto \bigl(x^{-\ell-k} e^{-x^2/4}\bigr)^{m_k+1} x^{m_k(m_k+1)/2}$, resulting from standard properties of Wronskians \cite{muir}, leads to
\begin{equation}
  \tilde{V}^{(2)}(x) = V_{\ell+k-m_k-1}(x) + m_k + 1 = \bar{V}^{(1)}(x) + m_k + 1,  \label{eq:V-tilde-2-bis}
\end{equation}
which establishes the desired relation between $V^{(1)}(x)$ and $\bar{V}^{(1)}(x)$. Since type II seed functions lead to isospectral transformations, the corresponding energies satisfy the relation
\begin{equation}
  \tilde{E}^{(2)}_{\ell+k-m_k-1,\nu} = \tilde{E}^{(1)}_{\ell+k,\nu} = E^{(1)}_{\ell+k,\nu} = 2\nu + \ell + k + 
  \frac{3}{2}, \qquad \nu=0, 1, 2, \ldots.
\end{equation}
We conclude that this third transformation allows us to go from $\tilde{H}^{(1)} = H^{(1)}$ to $\tilde{H}^{(2)} = \bar{H}^{(1)} + m_k + 1$.\par
%
%
We can now go from $H^{(2)}$ to $H^{(2)} + 2m_k + 2$ along the following path:
\begin{equation}
  \xymatrixcolsep{1.5pc}\xymatrix@1{
  H^{(2)}  \ar@/_{10mm}/[rrr]^{c}  \ar[r]^-{{\cal A}^{\dagger}} & H^{(1)}=\tilde{H}^{(1)}  
  \ar[r]^-{\tilde{\cal A}}  & \tilde{H}^{(2)} =
  \bar{H}^{(1)}+m_{k}+1 \ar[r]^-{\bar{\cal A}}  & \bar{H}^{(2)}+m_{k}+1=H^{(2)}+2m_{k}+2}
\end{equation}
which shows that the $(2m_k+2)$th-order operator
\begin{equation}
  c = \bar{\cal A} \tilde{\cal A} {\cal A}^{\dagger} 
\end{equation}
is a lowering operator for $H^{(2)}$.\par
%
%
Together with its Hermitian conjugate and $H^{(2)}$, it satisfies a PHA of $(2m_k+1)$th order, similar to (\ref{eq:PHA}) except that $Q(H^{(2)})$ is now given by
\begin{equation}
\begin{split}
  Q(H^{(2)}) &= \left(\prod_{i=1}^k (H^{(2)} - \alpha + 2m_i - k + 1)\right) \left(\prod_{j=0}^{m_k} 
       (H^{(2)} + \alpha - 2j + k - 1)\right) \\
  & \quad \times \left(\prod_{\substack{n=1 \\ n \ne m_k-m_{k-1}, \ldots, m_k-m_1}}^{m_k} (H^{(2)} - \alpha 
       - 2n - k -1)\right).  
\end{split}  
\end{equation}  
The action of $c$ and $c^{\dagger}$ on the wavefunctions of $H^{(2)}$ is not substantially different from that obtained in the harmonic oscillator case, so that the present PHA has $m_k+1$ infinite-dimensional unirreps too. Hence, the $k$ lowest states are mixed with the higher ones again.\par
%
%
\section{Conclusion}

In the present contribution, it has been shown that for the multi-step extensions of the harmonic oscillator and of the radial harmonic oscillator connected with type III EOP, it is possible to build new ladder operators $c$, $c^{\dagger}$ that satisfy a PHA with only infinite-dimensional unirreps. These new ladder operators differ from the usual ones $b$, $b^{\dagger}$, well known in SUSYQM. This has been made possible by combining the state-adding (or Darboux-Crum) and state-deleting (or Krein-Adler) approaches to the construction of these extensions.
Such new ladder operators are very useful to construct integrals of motion for superintegrable systems based on one-dimensional multi-step extensions connected with type III EOP.\par
%
%
An interesting open question is whether such a construction is possible for other types of potentials than the harmonic oscillator and the radial harmonic oscillator, such as potentials connected with Jacobi type III EOP.\par
%
%

\end{document}